\documentclass{article}

\usepackage{microtype}
\usepackage{graphicx}
\usepackage{subfigure}
\usepackage{booktabs} 

\usepackage{hyperref}

\usepackage[accepted]{mlsys2024}

\mlsystitlerunning{\smartfhe: Accurate Low-degree Polynomial Approximation of Non-Polynomial Operators for Fast Private Inference }

\usepackage{mathptmx} 

\usepackage{graphicx}%
\usepackage{multirow}%
\usepackage{amsthm}%
\usepackage{mathrsfs}%
\usepackage[title]{appendix}%
\usepackage{xcolor}%
\usepackage{textcomp}%
\usepackage{manyfoot}%
\usepackage{booktabs}%
\usepackage{listings}%
\usepackage{pifont}
\usepackage{comment}
\usepackage{xspace}
\usepackage{makecell}
\usepackage{subcaption}
\usepackage{amssymb,amsmath,amsthm,enumitem,amsfonts}
\usepackage[nodisplayskipstretch]{setspace}
\usepackage{colortbl}
\usepackage{booktabs}

\newcommand{\insertFigure}[2]{
    \begin{figure}[t!]
        \centering
        \includegraphics[width=\linewidth]{fig/#1.pdf}
	\vspace{-6mm}
        \caption{ #2}
	\vspace{-2mm}
        \label{fig:#1}
    \end{figure}
}

\newcommand{\insertFigureRatio}[3]{
    \begin{figure}[t!]
        \centering
        \includegraphics[width=#3\linewidth]{fig/#1.pdf}
        \caption{ #2}
	\vspace{-2mm}
        \label{fig:#1}
    \end{figure}
}

\newcommand{\insertWideFigureRatio}[3]{
    \begin{figure*}[ht!]
        \centering
        \includegraphics[width=#3\textwidth]{fig/#1.pdf}
	\vspace{-3mm}
        \caption{ #2}
	\vspace{-6mm}
        \label{fig:#1}
    \end{figure*}
}

\newcommand{\squishlist}{
 \begin{list}{$\bullet$}
  { \setlength{\itemsep}{0pt}
     \setlength{\parsep}{0pt}
     \setlength{\topsep}{3pt}
     \setlength{\partopsep}{0pt}
     \setlength{\leftmargin}{1.5em}
     \setlength{\labelwidth}{1em}
     \setlength{\labelsep}{0.5em} } }

\newcommand{\squishnums}{
 \begin{list}{$\bullets$}
  { \setlength{\itemsep}{0pt}
     \setlength{\parsep}{3pt}
     \setlength{\topsep}{3pt}
     \setlength{\partopsep}{0pt}
     \setlength{\leftmargin}{1.5em}
     \setlength{\labelwidth}{1em}
     \setlength{\labelsep}{0.5em} } }

\newcommand{\squishlisttwo}{
 \begin{list}{$\bullet$}
  { \setlength{\itemsep}{0pt}
     \setlength{\parsep}{0pt}
    \setlength{\topsep}{0pt}
    \setlength{\partopsep}{0pt}
    \setlength{\leftmargin}{2em}
    \setlength{\labelwidth}{1.5em}
    \setlength{\labelsep}{0.5em} } }

\newcommand{\squishend}{
  \end{list}  }

\newcommand{\smartfhe}{{\textbf{\texttt{SMART-PAF}}}\xspace}
\newcommand{\cmark}{\ding{51}}%
\newcommand{\xmark}{\ding{55}}%

\newcommand{\secref}[1]{\S\ref{#1}}
\newcommand{\figref}[1]{Fig.~\ref{#1}}
\newcommand{\tabref}[1]{Tab.~\ref{#1}}
\newcommand{\equref}[1]{Eq.~\ref{#1}}

\theoremstyle{thmstyleone}%
\theoremstyle{thmstyletwo}%

\theoremstyle{thmstylethree}%

\expandafter\def\expandafter\normalsize\expandafter{%
    \normalsize%
    \setlength\abovedisplayskip{-12pt}%
    \setlength\belowdisplayskip{4pt}%
    \setlength\abovedisplayshortskip{-2pt}%
    \setlength\belowdisplayshortskip{2pt}%
}

\newif\ifcommenton
\commentontrue

\ifcommenton
\newcommand{\TODO}[1]{\textcolor{red}{[TODO] #1}}
\newcommand{\JT}[1]{{\color{brown}\bfseries [Jianming: #1]}}
\newcommand{\AI}[1]{{\color{blue}\bfseries [Anirudh: #1]}}
\newcommand{\PC}[1]{{\color{blue}\bfseries [Prasanth: #1]}}
\newcommand{\TK}[1]{{\color{violet}\bfseries [TK: #1]}}
\newcommand{\fixme}[1]{{{\color{blue} #1}}}
\else
\newcommand{\TODO}[1]{}
\newcommand{\AI}[1]{}
\newcommand{\PC}[1]{}
\newcommand{\JT}[1]{}
\newcommand{\TK}[1]{}
\newcommand{\fixme}[1]{}
\fi

\begin{document}

\twocolumn[
\mlsystitle{\large Accurate Low-degree Polynomial Approximation of Non-Polynomial Operators for Fast Private Inference in Homomorphic Encryption}



\mlsyssetsymbol{equal}{*}
\vspace{-2mm}
\begin{mlsysauthorlist}
\mlsysauthor{Jianming Tong}{equal,goo}
\mlsysauthor{Jingtian Dang}{equal,to}
\mlsysauthor{Anupam Golder}{ed}
\mlsysauthor{Arijit Raychowdhury}{ed}
\mlsysauthor{Cong Hao}{ed}
\mlsysauthor{Tushar Krishna}{ed}
\end{mlsysauthorlist}
\vspace{-2mm}
\mlsysaffiliation{to}{Electrical and Electronics Engineering, Carnegie Mellon University, 5000 Forbes Avenue, Pittsburgh, 15213, Pennsylvania, U.S.}
\mlsysaffiliation{goo}{School of Computer Science, Georgia Institute of Technology, North Avenue, Atlanta, 30332, Georgia, U.S.}
\mlsysaffiliation{ed}{School of Electrical and Computer Engineering, Georgia Institute of Technology, North Avenue, Atlanta, 30332, Georgia, U.S.}

\mlsyscorrespondingauthor{Jingtian Dang}{dangjingtian@cmu.edu}
\mlsyscorrespondingauthor{Jianming Tong}{jianming.tong@gatech.edu}

\mlsyskeywords{Machine Learning, MLSys}

\vskip 0.3in

\begin{abstract}
As machine learning (ML) permeates fields like healthcare, facial recognition, and blockchain, the need to protect sensitive data intensifies. Fully Homomorphic Encryption (FHE) allows inference on encrypted data, preserving the privacy of both data and the ML model. 
However, it slows down non-secure inference by up to five magnitudes, with a root cause of replacing non-polynomial operators (ReLU and MaxPooling) with high-degree Polynomial Approximated Function (PAF).
We propose \smartfhe, a framework to replace non-polynomial operators with \textit{low-degree PAF} and then recover the accuracy of PAF-approximated model through four techniques: (1) Coefficient Tuning (CT) -- adjust PAF coefficients based on the input distributions before training, (2) Progressive Approximation (PA) -- progressively replace one non-polynomial operator at a time followed by a fine-tuning, (3) Alternate Training (AT) -- alternate the training between PAFs and other linear operators in the decoupled manner, and (4) Dynamic Scale (DS) / Static Scale (SS) -- dynamically scale PAF input value within $(-1, 1)$ in training, and fix the scale as the running max value in FHE deployment.
The synergistic effect of CT, PA, AT, and DS/SS enables \smartfhe to enhance the accuracy of the various models approximated by PAFs with various low degrees under multiple datasets. 
For ResNet-18 under ImageNet-1k, the Pareto-frontier spotted by \smartfhe in latency-accuracy tradeoff space achieves $1.42\times\sim 13.64\times$ accuracy improvement and $6.79\times \sim 14.9\times$ speedup than prior works. Further, \smartfhe enables a 14-degree PAF ($f_1^2\circ g_1^2$) to achieve $7.81\times$ speedup compared to the 27-degree PAF obtained by minimax approximation with the same 69.4\% post-replacement accuracy. Our code is available at \url{https://github.com/EfficientFHE/SmartPAF}
\end{abstract}
]



\printAffiliationsAndNotice{\mlsysEqualContribution} 

\section{Introduction}
\label{sec:intro}


As ML becomes more pervasive in fields such as healthcare~\cite{mateen2020improving}, facial recognition~\cite{facial_recog}, and blockchain~\cite{pipezk}, concerns regarding privacy leakage of private and sensitive data have arisen. Fully Homomorphic Encryption (FHE)~\cite{homomorphicStandard} provides a solution to these concerns by allowing for ML inference on encrypted data while preserving the privacy of both the data and models. However, FHE-based ML inference comes with a significant latency overhead, i.e. five orders of magnitude longer than the corresponding non-secure version~\cite{f1}. This slowdown is primarily due to non-polynomial operators (e.g. ReLU, MaxPooling, and Softmax, etc.), which dominate approximately half of the total latency~\cite{park2022aespa}. Therefore, a major research problem is \textit{how to efficiently process non-polynomial operators in FHE}.

\subsection{Challenges}
Non-polynomial operators pose a challenge in FHE due to the lack of native support. To overcome this, previous research has explored (1) a hybrid scheme, 
which offloads non-polynomial operators to other secure schemes with a secure data transfer to communicate data among schemes, and (2) approximation, which replaces non-polynomial operators with  polynomial approximation functions (PAF). 

The hybrid scheme is challenging in practice because a plethora of prior arts illustrate the prohibitive communication overheads for transferring data securely among different schemes~\cite{cryptonets, safenet, ran2022cryptogcn}.

\insertFigure{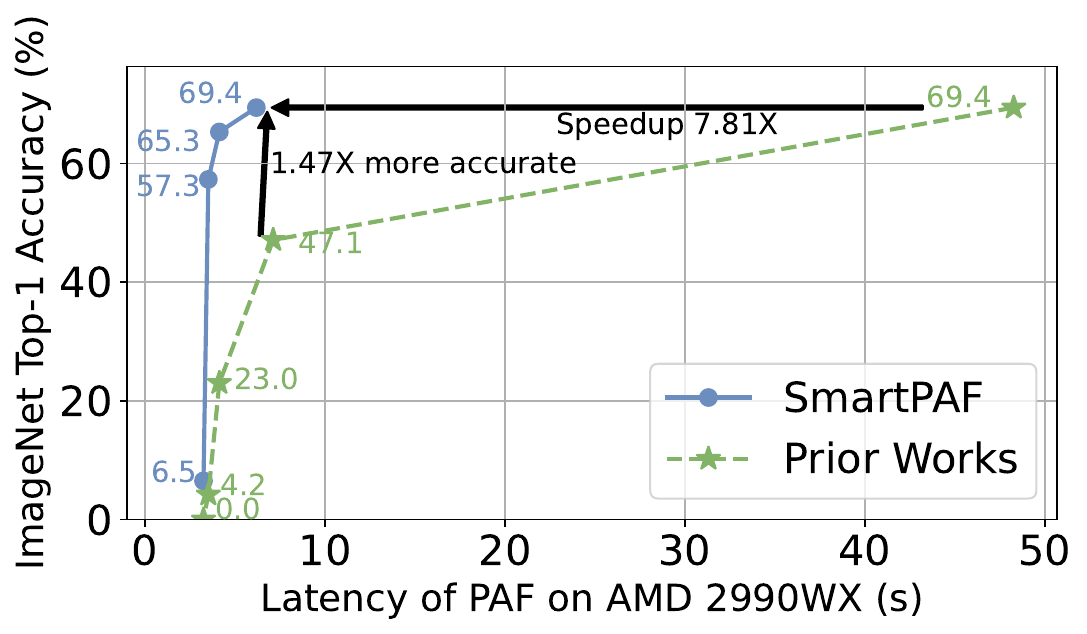}{\smartfhe replaces ReLU and MaxPooling by low-degree PAFs, and achieves better accuracy-latency Pareto-frontier than prior works~\cite{Lee2021PreciseAO,Minimax_approximation,cheon2020efficient} on ResNet-18 (ImageNet-1k).}

Despite its promise, PAF approximation is non-trivial because it requires striking a balance between accuracy and latency. A high-accuracy approximation requires high-degree PAFs with a prohibitively long chain of multiplications with bootstrapping. An example includes a SotA 27-degree PAF~\cite{Lee2021PreciseAO,BTS}. On the other hand, a low-latency approximation suffers from severe accuracy degradation, allowing only a subset of non-polynomial operators to be replaced with PAFs. In such cases, there are still some non-polynomial operators being offloaded to other schemes to reduce the accuracy drop, resulting in dominating communication latency overheads in practice~\cite{delphi, safenet, ran2022cryptogcn}. These approaches are suboptimal and highlight the need for exploring alternative approaches.


To address this challenge, previous research has explored coefficients fine-tuning to reduce accuracy drop when replacing non-polynomial operators with low-latency low-degree 
PAFs. 
However, existing techniques fail to converge for PAFs with degrees higher than 5, and PAFs with degrees lower than 5 still suffer from severe accuracy degradation for a simple 7-layer CNN model under CiFar-10 dataset~\cite{safenet}, let alone deeper ResNet-18 model under more complex task like ImageNet-1k. 
To explore the optimal degree with minimal accuracy degradation for various models under different tasks, the key challenge is to improve training techniques to enable the convergence of PAF with arbitrary degrees.  


%

\subsection{Our Contributions}
To tackle the aforementioned dilemma, this paper proposes four key techniques and a framework to enable the convergence of the PAF-approximated model when using PAFs with arbitrary degrees. This is the first-ever framework, to the best of our knowledge, to (1) replace all non-polynomial operators with $8\sim14$-degree PAFs, and (2) to adopt proposed training techniques to minimize accuracy degradation. Out contributions are listed as follows. (The comparison with prior arts is shown in \tabref{tab:cmp}) 

\textbf{Pre-Training Novel Techniques:}
\squishlist
\item \textbf{Coefficient Tuning (CT):} Using a uniform PAF to replace all non-polynomial layers neglects variations in input distributions, causing a marked drop in accuracy of PAF-approximated model (\tabref{tab:ablation}). CT refines PAF coefficients based on local input distributions, enhancing the accuracy of PAF-approximated model without fine-tuning by $1.04\times \sim 2.38\times$ across PAFs with various degrees.
\item \textbf{Progressive Approximation (PA):} Directly replacing all non-polynomial operators and training the full network, as adopted by previous works, lacks a theoretical convergence guarantee under SGD (\figref{fig:convergence_curve}).
Instead, PA sequentially replaces non-polynomial operators in inference order, with each substitution followed by training layers preceding the replacement point. The theoretical convergence of PA under SGD is analyzed in \secref{sec:PAF_Statement}, resulting in an accuracy boost of $0.4\%\sim 1.9\%$.
\squishend

\textbf{In-Training Novel Techniques:}
\squishlist
\item \textbf{Alternate Training (AT):} When training PAF and linear operator (like convolution) in a PAF-approximated model together, issues arise: slow convergence with a small learning rate or divergence with a large one. 
This stems from the different hyperparameter needs of PAF coefficients and linear layers training.
Therefore, AT trains PAF coefficients and other layers separately, using different hyperparameters, and alternates between PAF coefficients training and other layers training. This leads to accuracy climbing by $0.6\%\sim2.3\%$ (\tabref{tab:ablation}).
\item \textbf{Dynamic Scaling (DS) and Static Scaling (SS):} PAF training is sensitive to the input value range. Prior works take a fixed scale value to reduce the input range in training because FHE does not support value-dependent operators. However, input values in training might overflow under a fixed scale, hampering fine-tuning accuracy and potentially leading to divergence. We address this by introducing Dynamic Scaling, normalizing inputs to the $[-1, 1]$ range during training. Subsequently, we apply Static Scaling to the post-training model, anchoring the scale to the peak value in FHE deployment. 
\squishend
\textbf{\smartfhe Framework:}
\squishlist
\item The order of applying proposed techniques affects final accuracy, and thus a systematic scheduling framework is proposed to automatically perform non-polynomial replacement and PAF-approximated model training. Our evaluation results show that for ResNet-18 (ImageNet-1k) and VGG-19 (CiFar-10), \smartfhe consistently enables low-degree PAFs to demonstrate higher accuracy than high-degree SotA PAFs (\tabref{tab:ablation} and \tabref{tab:compare_vgg_cifar10}). For example, \smartfhe identifies the Pareto-frontier of various PAFs shown in \figref{fig:pareto_frontier_cmp} and spots the sweet-point PAF with 14 degrees, which can achieve the same 69.4\% validation accuracy with $7.81\times$ latency speedup in ResNet-18 inference on ImageNet-1k, compared to the minimax approximation with 27-degree PAF~\cite{Lee2021PreciseAO}.
\squishend

This work pushes the state-of-the-art in PAF-approximated model training for higher accuracy with lower latency (\figref{fig:pareto_frontier_cmp}). First, we formalize PAF-approximated model training into an optimization problem (\secref{sec:PAF_Statement}). Second, \secref{sec:proposedmethod} elaborates on the \smartfhe techniques and framework with both intuitive and theoretical analysis. Third, we individually assess the \smartfhe techniques for VGG-19 (CiFar-10) and ResNet-18 (ImageNet-1k) in \secref{sec:exp}.

\begin{table}[!t]\centering
\caption{Comparison of \smartfhe with prior works~\cite{safenet, ran2022cryptogcn, cryptonets, Hesamifard2017CryptoDLDN, Lee2021PreciseAO, CHE, f1, BTS, CraterLake, cheetah, Gazelle, delphi, heax, SHE}.}\label{tab:cmp}
\scriptsize
\resizebox{0.48\textwidth}{!}{
\begin{tabular}{ccccc}\hline
&\makecell{Low \\ Communication \\ Overhead} &\makecell[c]{Low \\ Accuracy\\ Degradation} & \makecell[c]{Low\\ Latency \\ Overhead} \\\hline
SafeNet, CryptoGCN &\xmark &\xmark & \cmark \\
CryptoNet, CryptoDL, LoLa, CHE &\xmark &\xmark &\cmark  \\
F1, CraterLake, BTS &\cmark  &\cmark  &\xmark \\
HEAX, Delphi, Gazelle, Cheetah &\xmark &\xmark  &\cmark \\
SHE &\cmark  &\cmark  &\xmark \\
\smartfhe &\cmark  &\cmark  &\cmark  \\
\hline
\end{tabular}}
\end{table}

\section{Background}
\label{sec:back}
\subsection{Non-polynomial Operators in FHE-based ML Inference}

\insertFigure{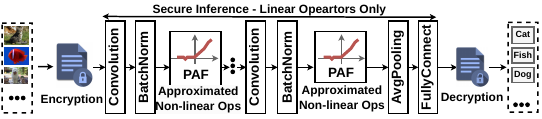}{Overview of the FHE-base ML inference where original non-polynomial operators are replaced by Polynomial Approximated Activation (PAF). \vspace{-4mm}}

FHE is an asymmetric encryption scheme that enables ciphertext-based computation with Cheon-Kin-Kim-Song (CKKS)~\cite{CKKS} as the most commonly used FHE scheme for machine learning inference due to its superior efficiency in approximate computation compared to other schemes such as BGV, BFV, and TFHE~\cite{heax}. Under the CKKS scheme, only polynomial operators are allowed as shown in \figref{fig:goal} such that all non-polynomial operators including both ReLU and MaxPooling must be replaced by PAF (Approximation) or offloaded to other schemes (Hybrid Scheme). Prior works have also shown that approximation offers superior performance and lower overhead compared to the hybrid scheme~\cite{safenet}.

\subsection{Polynomial Approximated Function (PAF)}
High-degree polynomials are theoretically capable of approximating arbitrary functions, but approximating ReLU or MaxPooling directly can result in severe approximation errors~\cite{Minimax_approximation}. Prior research~\cite{Minimax_approximation} has shown that it is more effective to approximate the $sign(x)$ function~\footnote{$sign(x)$ outputs $1/-1$ if $x$ is positive/negative, and $0$ for zero} and then construct the ReLU and Max operator using it, such as with $\frac{(x+sign(x)\cdot x)}{2}$ and $\frac{(x+y) + (x-y)\cdot sign(x-y)}{2}$.

Lower latency and higher accuracy (low accuracy degradation) are two fundamental goals of replacing non-polynomial operators with PAFs. 
Latency is determined by the degree of the polynomial, as FHE-based multiplication dominates latency.
Accuracy degradation arises from the difference between the PAF and the original non-polynomial operators, as PAF cannot precisely approximate targeted non-polynomial functions. Such differences can vary depending on the degree of the PAF, the cascaded format of the PAF, and the coefficient values of the PAF.

We adopt cascaded polynomial based PAFs throughout the paper because they achieve lower approximation error than a single polynomial with the same degree~\cite{Minimax_approximation, Lee2021PreciseAO}. 
\tabref{tab:PAF_baseline} shows the selected PAFs with the minimal multiplication depth under different degrees constraints. 
We use the notation $f^n$ to indicate a serial nested function call of the same polynomial $f$ for $n$ times, for example, $f^2=f(f(x))$.

\subsection{PAF Approximation Input Range}
\label{sec:paf_input_range}
The effectiveness of PAFs in approximating non-polynomial operators depends on their input range. 
For example, setting the approximation input range as $[-1, 1]$ for a PAF indicates that it gives relatively more accurate approximations of non-polynomial operators' output when the input falls in $[-1, 1]$ than outside this range. 

However, determining an appropriate input range can be challenging because a narrow input range can lead to severe approximation loss for input value outside the input range and potentially lead to training divergence. In contrast, a broader range may result in a large average approximation error across the entire input range, because of the limited representation capability of PAFs to capture negligible differences among input values under a broad range. 

Prior works~\cite{Lee2021PreciseAO, safenet} adopted a fixed scale in training because FHE does not support value-dependent operation. Specifically, ~\cite{Lee2021PreciseAO} determines the fixed scale by adding some margin to the maximum of input values. However, input values in training might overflow the range specified by a fixed scale, hampering fine-tuning accuracy and potentially leading to divergence. Therefore, we propose dynamic-scale training and only pick a fixed scale in model deployment.

\begin{table}[!t]\centering
\caption{PAF and corresponding multiplication depth. $\alpha$ indicates precision parameter~\cite{Lee2021PreciseAO}, and $f_i, g_i$ refer to PAF base in \cite{cheon2020efficient} }\label{tab:PAF_baseline}
\scriptsize
\resizebox{0.48\textwidth}{!}{
\begin{tabular}{ccccccccccccccc}\hline
Form & $\alpha=10$ & $f_1^2 \circ g_1^2$ & $\alpha=7$ & $f_2\circ g_3$ & $f_2\circ g_2$ & $f_1\circ g_2$ \\\hline
Degree & 27 & 14 &12 &12 &10 &5 \\
Multiplication Depth & 10 & 8 & 6 & 6 & 6 & 5 \\
\hline
\end{tabular}}
\vspace{-4mm}
\end{table}

\section{Problem Statement}
\label{sec:problem_statement}
\subsection{Polynomial Approximation Problem Statement}
\label{sec:PAF_Statement}
The essential problem for approximating non-polynomial operators with 
PAF is the regression problem shown in \equref{equ:non_convex}. Specifically, the goal is to find a vector of coefficients $\mathbf{a}_i = \{a_i^0, \cdots, a_i^N\}$ for a N-degree PAF, such that the cumulative error of replacing non-polynomial function $R(\mathbf{x}_i)$ by a $N$-degree PAF is minimal given input data as $\mathbf{x}_i\ (i=0,\cdots, D-1)$, where $D$ stands for the total number of non-polynomial layers \footnote{we use non-polynomial layers and non-polynomial operators interchangeably.}~\cite{boyd2004convex}. 

\begin{equation}
\label{equ:non_convex}
\resizebox{0.91\hsize}{!}{
    $\min\limits_{\mathbf{a}_0,\cdots,\mathbf{a}_{D-1}} f(\mathbf{a},\mathbf{x}) =$ \\
    $\sum_{i=0}^{D-1} \left[ \frac{1}{N} \sum_{i=0}^{N-1} ( R(\mathbf{x}_i, \mathbf{a}_0,\cdots, \mathbf{a}_i) -  \mathbf{a}_i\cdot \mathbf{x}_i )^2 \right]$ }
\end{equation}
where $\mathbf{x}_i = \{x_i^0, x_i^1, \cdots, x_i^N\}$. The subscript / superscript of $x$ and $a$ indicates layer index / degree index. On the other hand, the subscript / superscript for $\mathbf{a}$ and $\mathbf{x}$ indicates layer index / the training epoch. After replacing $0, \cdots, (i-2)$-th ReLU by PAFs, the input of the $i$-th ReLU $\mathbf{x}_i$ becomes dependent on $\mathbf{a}_0, \cdots, \mathbf{a}_{i-1}$, i.e. $i$-th ReLU becomes $R(\mathbf{x}_i, \mathbf{a}_0,\cdots, \mathbf{a}_i)$ instead of $R(\mathbf{x}_i)$. 

The \equref{equ:non_convex} is non-convex as it optimizes all $D$ vectors of coefficients $\mathbf{a_i},\ i \in [0,D)$ under the changing $\mathbf{x}$.

\begin{equation}
\label{equ:regression}
    \min\limits_{\mathbf{a}_i} f'(\mathbf{a}_i) = \\
    \frac{1}{N} \sum_{i=0}^{N-1} ( R(\mathbf{x}_i) -  \mathbf{a}_i\cdot \mathbf{x}_i )^2
\end{equation}

The \equref{equ:regression} only optimizes a single layer, such that $\mathbf{x}_i$ statistically is fixed, which makes it convex. In such cases, the stochastic gradient descent could guarantee finding the optimal solution $\mathbf{a}_i^*$ given initial value as $\mathbf{a}_i^0$. The error could converge in the speed order as shown in \equref{equ:error_decrease} after training $i$-th non-polynomial operator coefficients $\mathbf{a}_i$ for $T$ epochs~\cite{boyd2004convex} $i \in [0,D)$. And a good initialization value of $\mathbf{a}_i^0$ could reduce the total error and thus improve the convergence speed.

\begin{equation}
    \label{equ:error_decrease}
    f'(\mathbf{a}_i^T) - f'(\mathbf{a}_i^*) = O(\frac{||\mathbf{a}_i^0 - \mathbf{a}_i^*||^2}{\sqrt{T}})
\end{equation}

where $\mathbf{a}_i^T$ indicates the value of $\mathbf{a}_i$ after training for $T$ epochs.

\subsection{Post-replacement Model Retraining}
The PAF-approximated model replaces non-polynomial operators with PAFs. Such a replacement changes the original model structure, leading to changed data distribution of $x_i$ for all layers after the replacement points. Therefore, original coefficients trained using the original model structure may perform worse under new data distribution, leading to accuracy degradation. Therefore, the PAF replacement forces retraining to fine-tune coefficients to learn the changes in input data distributions.

\section{\smartfhe Techniques}
\label{sec:proposedmethod}
\subsection{Overview} 
Training methods adopted by prior arts lead to divergence when training PAF with a degree higher than 5, forbidding exploration using PAFs with higher degrees to enhance overall accuracy. In this section, we analyze critical reasons behind the divergence of existing training algorithms and propose corresponding four techniques to guarantee training convergence while replacing non-polynomial operators with PAFs of arbitrary degrees. Then we propose a framework to schedule techniques automatically.


\subsection{Coefficient Tuning (CT)}

\equref{equ:error_decrease} indicates that a good initialization can reduce the overall training time, facilitating faster convergence, which has been ignored by prior arts~\cite{Lee2021PreciseAO,cheon2020efficient,Minimax_approximation,safenet}. These studies have typically initialized PAFs with the same coefficients using traditional regression algorithms. Such initialization is suboptimal because it overlooks the distribution differences of data at different non-polynomial operators.

Therefore, we propose coefficient tuning (CT) as a technique to use profiled data distribution~\footnote{We assume that distribution probability range is consistent across input samples.} to obtain a closer-to-original initialization point and then use PAF with different initialization points to replace different non-polynomial operators, as shown in \figref{fig:CT}.  The CT involves the following steps: (1) obtain the coefficients of PAFs using traditional regression methods on the full range of input data; (2) profile the input data distribution for each non-polynomial operator; (3) tuned PAF coefficients to minimize overall approximation errors on profiled distributions to obtain the post-CT PAFs; (4) replace original non-polynomial operators with the post-CT PAFs.

Intuitively, CT tunes PAFs to achieve higher accuracy and less approximation error over a reduced smaller input range, which corresponds to the highest-probability range in the data distribution. 
Besides, CT reduces training time because the initial value of coefficients $\mathbf{a^0}$ are closer to the optimal value $\mathbf{a^*}$ as indicated by \equref{equ:error_decrease}.

\insertFigure{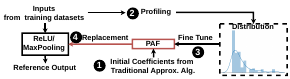}{Coefficient Tuning (CT) uses profiled distribution to tune PAF coefficients to generate more accurate results on a reduced smaller input range (high-probability range in profiled data distribution), leading to closer-to-optimal initialization, higher accuracy, and lower training time}

\subsection{Progressive Approximation (PA)}
\label{sec:PA}

Replacing all non-polynomial operators in the given ML model with PAFs simultaneously, as done by prior arts, is a suboptimal approach. This is because approximation errors of early replacements propagate to all layers behind the replacement point, making the regression (\equref{equ:regression}) non-convex by varying both PAF coefficients $\mathbf{a}_i$ and input data $\mathbf{x}_i$ ($i \in [0,D)$) for all $D$ non-polynomial layers, causing training divergence. Instead, we propose the Progressive Approximation (PA) approach, which replaces the non-polynomial layer
with PAF, one layer at a time, followed by fine-tuning of PAF coefficients until accuracy convergence as shown in \figref{fig:PA}. In such cases, all $\mathbf{x}_i$ ($i \in [0,D)$) statistically get fixed because we only vary PAF coefficients of a single layer, which is one of all ($\mathbf{a}_0, \mathbf{a}_1,\cdots, \mathbf{a}_{D-1}$), ensuring a simple convex regression problem shown in \equref{equ:regression}, which is easily optimizable by SGD.

For example in \figref{fig:PA}, PA starts with replacing the first ReLU with a PAF, and then fine-tunes PAF coefficients until accuracy converges. Then PA repeats the same flow for the following non-polynomial operators. 

Intuitively, the replacement of all non-polynomial operators by PAFs simultaneously introduces a huge deviation to the original model which is hard for the training algorithm to recover, while PA applies the overall approximation error progressively instead of applying all errors at once, restricting the approximation error to the optimizable range of the training algorithm to enable convergence and effectively mitigate the accuracy degradation. 

\insertFigure{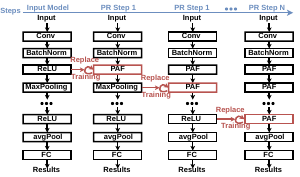}{Progressive Approximation (PA) progressively replaces non-polynomial operators, one layer at a time followed by coefficients fine-tuning, to guarantee an SGD-optimizable convex regression problem shown in \equref{equ:regression}, enabling training convergence for replacing the targeted model with PAFs of arbitrary degrees.}

\subsection{Alternate Training (AT)}
Replacing the non-polynomial operators with PAFs requires model fine-tuning, as pretrained model parameters are no longer optimal for the new PAF-approximated model. 
However, multiple previous works fine-tune the PAF-approximated models as a whole to optimize both PAF coefficients and the parameters of other layers (like Convolution, BatchNorm, etc.). Such a training scheme, however, deteriorates accuracy, indicating a failure of convergence. We demonstrate it later in \figref{fig:convergence_curve}. 
This is because modifications to PAF coefficients can have a significant impact on final inference results, unlike changes to convolution weights, which may not affect overall results at all. Such an observation is consistent with the model structure that all data need to go through PAF while some weights can be pruned without any impact on overall accuracy.

To tackle training divergence, PAF coefficients fine-tuning should be decoupled from fine-tuning parameters of other layers, and two fine-tuning processes should use different training hyperparameters. We thus propose Alternate Training (AT) to fine-tune PAF coefficients and parameters of other layers separately in an alternate manner.
Specifically, AT fine-tunes PAF coefficients first while keeping other parameters fixed, as shown in AT step 1 in \figref{fig:AT}.
After reaching a specific epoch threshold, PAF coefficients get frozen and AT fine-tunes parameters of other layers in AT step 2. Note that the training in different AT steps may use different hyperparameters because of different parameter sensitivity in different layers. AT repeats until the accuracy finally converges, often resulting in an accuracy climb. 

Intuitively, AT considers the sensitivity difference between PAFs parameters and parameters of linear layers, and decouples the two training processes to avoid training interference. 

\insertFigure{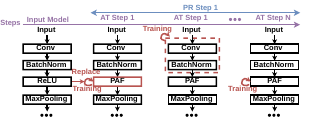}{Alternate Training (AT) fine-tunes PAF coefficients and parameters of other layers separately in an alternate manner.}

\subsection{Dynamic Scaling (DS) and Static Scaling (SS)}

In Section \ref{sec:paf_input_range}, we discuss the challenge of setting an appropriate input range for PAFs. Previous studies have employed a large input range, such as $[-50, 50]$, to prevent infinite errors for values outside the range \cite{Lee2021PreciseAO}. However, we contend that this approach is suboptimal because it results in high approximation error across the entire input range, which can impede training convergence and lead to significant accuracy degradation (e.g., accuracy drops from 69\% to 10\% for ResNet-18 on ImageNet-1k).

To tackle this issue, we introduce Dynamic Scaling (DS) and add an auxiliary layer before each PAF to normalize the value range of its input data to $[-1, 1]$ during fine-tuning. DS determines the scale on a batch-by-batch basis. For each batch, the scale is set to the highest absolute input value, and then all data are proportionally scaled to fit within the $[-1, 1]$ range. This ensures that the inputs within each batch spread the $[-1, 1]$ range, maximizing their relative differences for better distinguishment while staying within bounds.

However, the post-training model is intended for FHE deployment, where Dynamic Scaling (DS) is inapplicable because FHE cannot select the batch's maximal value due to the absence of value-dependent operators. Therefore, we propose Static Scaling (SS) to fix the scale of each auxiliary layer in \textit{PAF-approximated model} as the input running maximum under the training dataset 
Such a scale determination relies on the similarity between training data and validation data. In our evaluation, either a higher or smaller scale results in lower accuracy for both VGG-19 (CiFar-10 dataset) and ResNet-18 (ImageNet-1k dataset).



\subsection{\smartfhe Framework}
\insertFigure{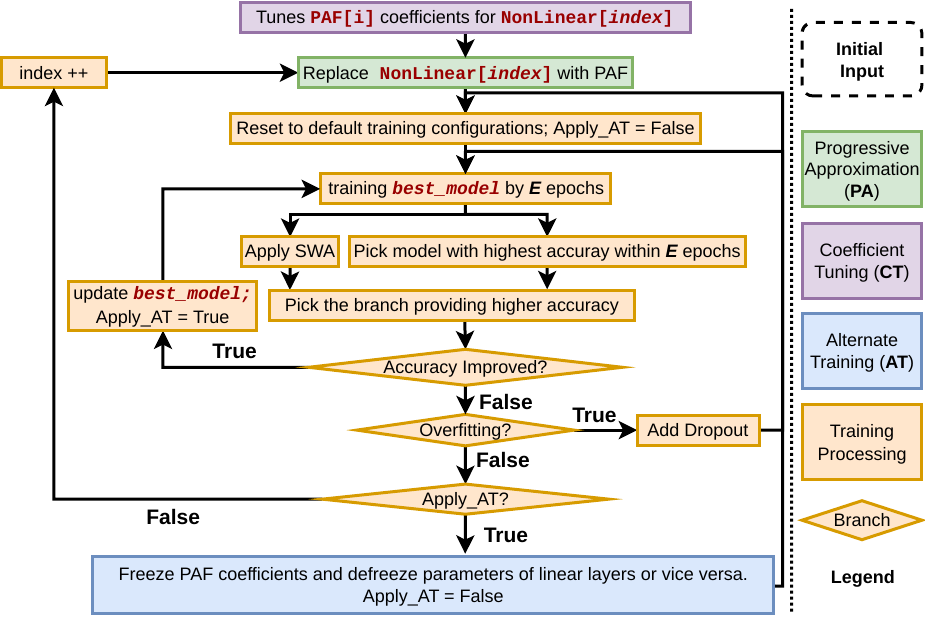}{Overview of \smartfhe framework, which automatically schedules \smartfhe techniques including CT, PA, AT, and existing techniques including Dropout and Stochastic Wegiths Averaging (SWA).}


The sequence in which CT, PA, AT, and DS/SS are applied can greatly affect the final validation accuracy. Furthermore, employing existing techniques, such as dropout for overfitting mitigation and Stochastic Weights Averaging (SWA) for faster convergence, is crucial.

To navigate these complexities, we introduce a scheduler in \smartfhe. This scheduler systematically applies training configurations and proposed techniques, organizing the training into steps, with each step replacing a single non-polynomial layer with PAF. Coefficient Tuning is applied offline before all steps. The sequence within a single step is outlined below:

\squishlist
\item \textit{Training Group:} A training group first trains the PAF-approximated model for $E$ epochs, followed by applying SWA using weights of all $E$ epochs. The model with the highest validation accuracy proceeds to the following procedures.
\item \textit{Accuracy Improvement Detection:} If the validation accuracy gets improved during the process, the framework launches a new training group until no accuracy improvement is observed.
\item \textit{Overfitting Avoidance:} The framework applies Dropout followed by a new training group if spots overfitting~\footnote{we adopt the empirical overfitting condition, which is ``training accuracy $>$ validation accuracy + $10\%$"}. 
\item \textit{Alternative Training:} When previous techniques do not improve accuracy anymore, AT is applied to swap training targets between PAF and other layers, followed by a new training group.
\item \textit{Step Termination Condition:} Current step terminates when observing no accuracy improvement in above processes. 
\squishend

\smartfhe framework dedicates one step for each non-polynomial layer, orchestrating these steps progressively based on the inference order (Progressive Approximation). Dynamic Scaling is adopted in the entire fine-tuning process, while Static Scaling is applied for the post-finetuning model.

\section{Evaluation}
\label{sec:exp}

\subsection{Evaluation Setup} 
\textbf{Model} We evaluated \smartfhe with models being used by prior works for fair comparison~\cite{Lee2021PreciseAO, safenet}, including VGG-19 (18 ReLU and 5 MaxPooling), and ResNet-18 (17 ReLU and 1 MaxPooling). 

\textbf{Datasets} We adopt CiFar-10 and ImageNet-1k~\cite{resnet18, krizhevsky2012imagenet} to test the generality of proposed techniques for tasks with different complexity.

\textbf{PAF Form} We adopt 6 PAFs (\tabref{tab:PAF_baseline}) with minimal multiplication depth under different degrees as the approximation of $sign(x)$. ReLU and Max operators are replaced by $\frac{(x+sign(x)\cdot x)}{2}$ and $\frac{(x+y) + (x-y)\cdot sign(x-y)}{2}$, separately.

\textbf{Latency Evaluation} We implement PAF in Microsoft SEAL library~\cite{SEAL} using CKKS~\cite{CKKS} (Degree: 32768, modulus bitwidth: 881) and evaluate the wall-clock PAF latency on AMD Threadripper 2990WX. 

\textbf{Accuracy Evaluation} The accuracy is obtained through evaluating PAF-approximated models under given datasets. 

\textbf{Training Hyperparameters} We adopt different training hyperparameters for training PAF coefficients and parameters of other layers, as shown in \tabref{tab:set_up}. We set the epoch $E=20$ for each training group. A smaller epoch leads to negligible accuracy changes while a larger epoch leads to long latency of SWA which slows down the training processing.  
\insertFigureRatio{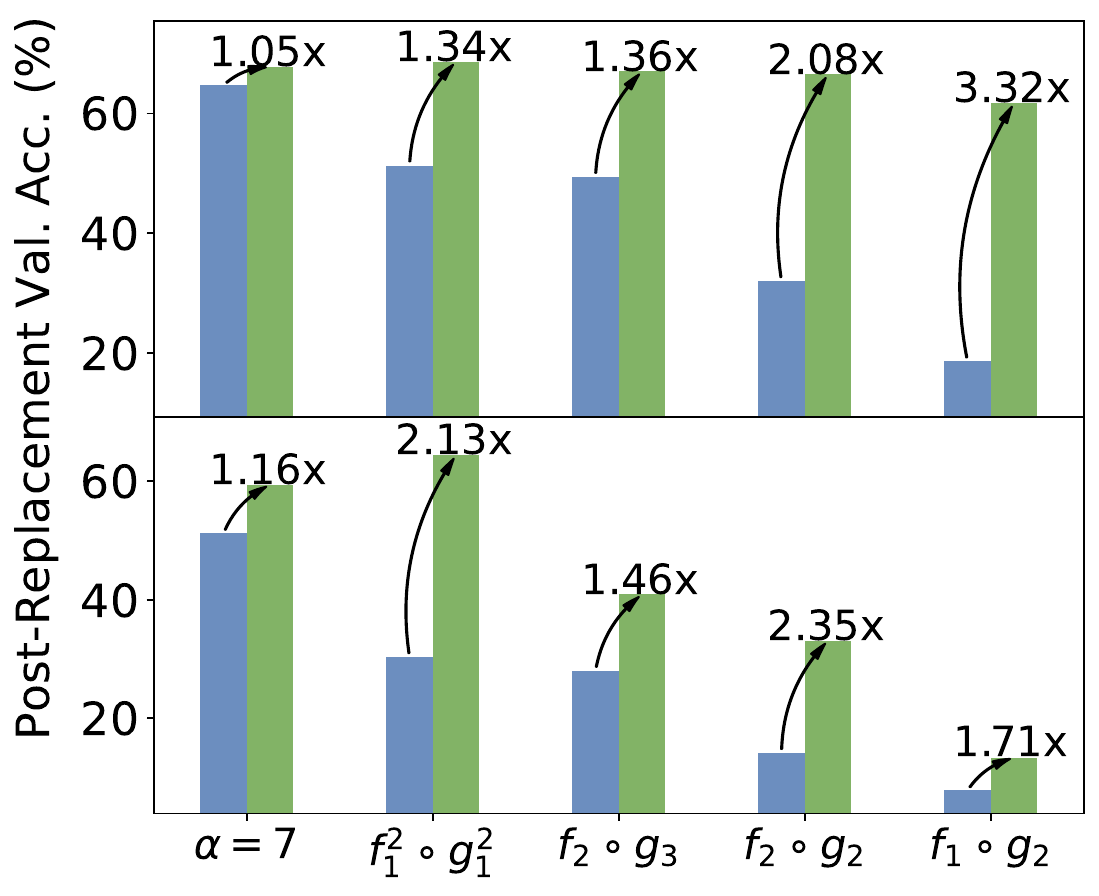}{Comparison of ResNet-18 (ImageNet-1k) validation accuracy w/o Fine-Tuning between Coefficient Tuning (CT, green bar) and baseline (blue bar). Top: only replace ReLU by PAF. Bottom: replace all ReLU and MaxPooling.}{0.9}

\subsection{Coefficient Tuning (CT) Evaluation}
CT adjusts PAF coefficients based on profiled value distributions and improve the $1.05\sim3.32\times$ post-replacement validation accuracy w/o fine tuning, as shown in \figref{fig:Coefficient_tuning}.

When only approximating ReLU, CT shows more benefits for polynomials with lower degrees, as they have less capability to fit the entire input range and are more prone to significant accuracy reduction. CT mitigates this loss by focusing on fitting the high-probability region in the distribution, resulting in less initial post-replacement approximated error. On the other hand, polynomials with higher degrees have less overall approximated error across the entire input range and, therefore, show less improvement from CT. 

Further, approximating both ReLU and MaxPooling leads to $10.9\%\sim21\%$ accuracy drop than approximating ReLU only, as shown by comparing the blue bar in the top and bottom figures in \figref{fig:Coefficient_tuning}. Contrary to the ReLU approximations in earlier works~\cite{park2022aespa}, this suggests that solely assessing ReLU does not provide a full picture of PAF's non-polynomial approximation prowess. 
Comprehensive evaluations should encompass both ReLU and MaxPooling, wherein CT consistently aids in accuracy restoration, improving up to $34.1\%$ accuracy for $f_1^2 \circ g_1^2$!


\insertFigure{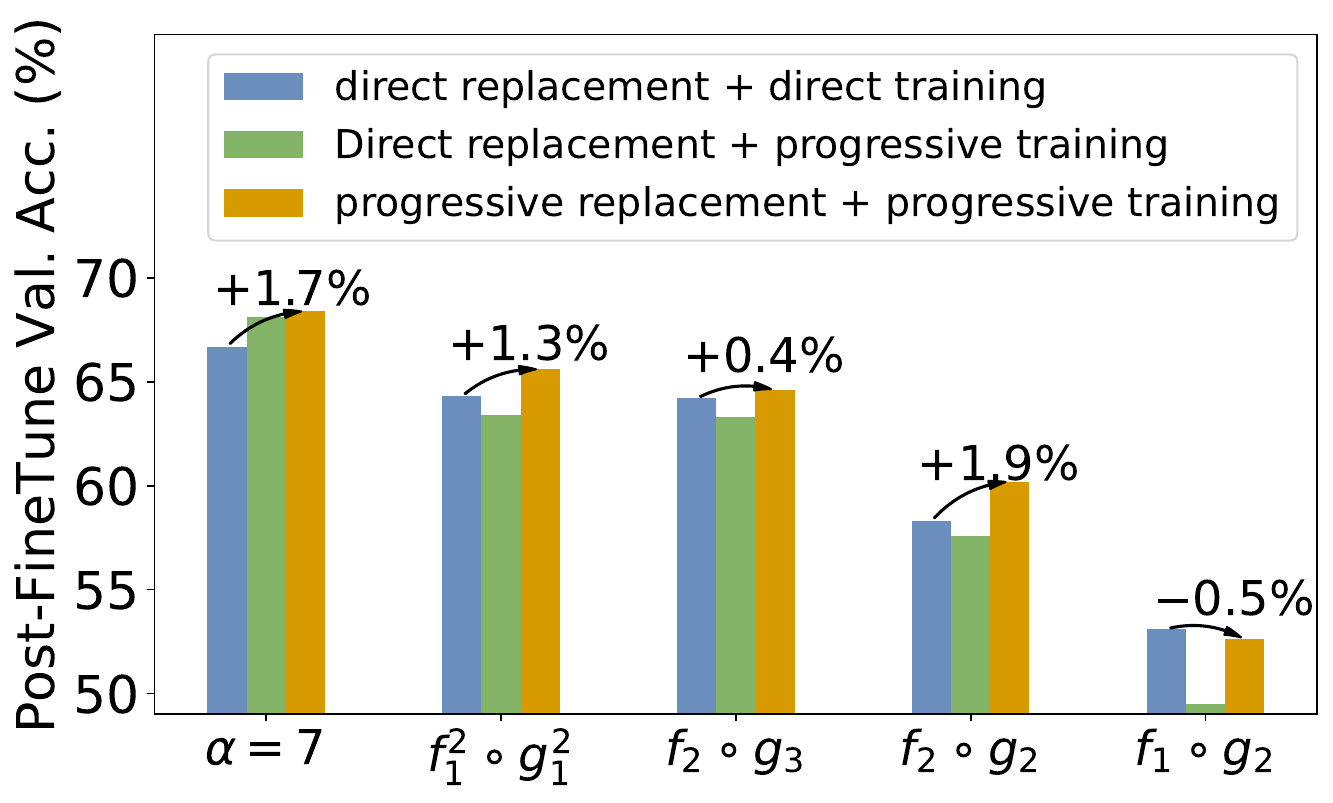}{Comparison of ResNet-18 (ImageNet-1k) validation accuracy w/ Fine-Tuning between Progressive Approximation (PA, orange) and baseline (blue). Green bar only adopts progressive training w/o progressive replacement and suffers from severe accuracy degradation.}

\vspace{-2mm}
\subsection{Progressive Approximation Evaluation}

During fine-tuning, the baseline (prior works) uses a direct replacement + direct training approach: it replaces all non-polynomial operators with PAFs and trains other layers, excluding the PAFs. In contrast, PA adopts a step-wise strategy. In each step, it replaces one non-polynomial operator with a PAF and trains preceding layers progressively (termed progressive replacement and progressive training). This method yields an accuracy improvement ranging from  $0.4\sim1.7$\%, as illustrated in \figref{fig:Progressive_Approximation}. Significantly, the progressive replacement is pivotal for this improvement. This is evident from the marked accuracy degradation of the green bar (direct replacement + progressive training) relative to the orange bar. This is because progressive replacement simplifies the optimization, enabling its theoretical convergence under SGD, as discussed in \secref{sec:PAF_Statement}.

\begin{table*}[!t]\centering
\caption{Ablation study under of ResNet-18/VGG-19 models under CiFar-10/ImageNet-1k datasets. CT: Coefficient Tuning; PA: Progressive Approximation; AT: Alternate Training; DS: Dynamic Scaling; SS: Static Scaling. DS is only used in fine-tuning to improve accuracy and must converted to SS to be adopted in Homomorphic Encryption (HE) because DS contains value-dependent operators that are not supported by HE. Best HE-compatible accuracies are colored in grey while results with the best validation accuracy are bolded. An accuracy of 0\% indicates training divergence.}
\vspace{-3mm}
\scriptsize
\begin{tabular}{cccccccc}\toprule
Model-Dataset& Technique Setup & $f_1^2 \circ g_1^2$ & $\alpha=7$ & $f_2\circ g_3$ & $f_2\circ g_2$ &$f_1\circ g_2$ \\\hline
\multirow{9}{*}{\makecell[c]{Replace ReLU \\ ResNet-18 \\ ImageNet-1k\\ Original Accuracy \\ $69.3$\%}} & baseline + DS w/o fine tune &51.30\% & 64.70\% &49.40\% &32.00\% &18.60\% \\
& baseline + CT + DS w/o fine tune &68.60\% &67.70\%  &67.00\% &66.50\% &61.70\% \\
& baseline + DS &64.30\% &66.70\% &64.20\% &58.30\% &53.10\% \\

& baseline + AT + DS &65.20\% &68.30\%  &63.70\% &60.50\% &52.00\% \\
& baseline + PA + DS &65.60\% & \textbf{68.40\%}  &64.60\% &60.20\% &52.60\% \\
& baseline + PA + AT + DS &64.90\% &67.40\% &64.60\% &56.50\% &47.10\% \\
& baseline + CT + PA + DS &68.20\% &67.00\% &\textbf{67.60\%} &65.90\% &60.80\% \\
&baseline + CT + PA + AT + DS & \textbf{69.00\%}  &68.10\% &61.40\% & \textbf{66.50\%} &\textbf{63.10\%} \\
& \makecell[c]{Accuracy Improvement over ``baseline + DS"} &+4.7\%(1.07$\times$) &+1.7\%(1.03$\times$) &+3.4\%(1.05$\times$) &+8.2\%(1.14$\times$) &+10\%(1.19$\times$) \\

\hline
\multirow{7}{*}{\makecell[c]{Replace all \\ non-polynomial \\ ResNet-18 \\ ImageNet-1k \\ Original Accuracy \\ $69.3$\%}} & baseline + DS w/o fine tune & 30.3\% & 51.2\% & 28\% & 14.1\% & 7.8\% \\
& baseline + CT + DS w/o fine tune & 64.4\% & 59.4\% & 40.9\% & 33.1\% & 13.3\% \\
& baseline + DS & 59.6\% & 66.2\% & 62\% & 49\% & 37\% \\
& baseline + SS (\textbf{prior work~\cite{Minimax_approximation}}) & 25.5\% & 47.1\% & 23\% & 4.2\% & 0\% \\
& baseline + CT + PA + AT + DS & \textbf{69.9\%} & \textbf{68\%} & \textbf{65.7\%} & \textbf{64.1\%} & \textbf{57.8}\% \\
& \smartfhe: baseline + CT + PA + AT + SS & \cellcolor[HTML]{d3d3d3}69.4\% & \cellcolor[HTML]{d3d3d3}67\% & \cellcolor[HTML]{d3d3d3}65.3\% & \cellcolor[HTML]{d3d3d3}57.3\% & \cellcolor[HTML]{d3d3d3}6.5\% \\
&\makecell[c]{Accuracy Improvement over~\cite{Minimax_approximation}} &+43.9\%(2.72$\times$) & +19.9\%(1.42$\times$) & +42.3\%(2.84$\times$) & +53.1\%(13.64$\times$) &+6.5\%($\infty$) \\
\hline
\multirow{6}{*}{\makecell[c]{Replace all \\ non-polynomial \\ VGG-19 \\ CiFar-10 \\ Original Accuracy \\ 93.95\%}} & baseline + DS & 93.4\% & 92.38\% & 89.87\% & 89.87\% & 86.57\% \\
& baseline + SS (\textbf{prior work~\cite{Minimax_approximation}}) & 91.06\% & 81.35\% & 76.58\% & 58.11\% & 43.84\% \\
& baseline + CT + DS & 93.39\% & 93.6\% & 93.3\% & \textbf{92.4\%} & \textbf{91.53\%} \\
& baseline + CT + PA + AT + DS & \textbf{93.6\%} & \textbf{93.81\%} & \textbf{93.59\%} & 91.49\% & 91.51\% \\
& \smartfhe: baseline + CT + PA + AT + SS & \cellcolor[HTML]{d3d3d3}92.16\% & \cellcolor[HTML]{d3d3d3}92.62\% & \cellcolor[HTML]{d3d3d3}91.51\% & \cellcolor[HTML]{d3d3d3}88.45\% & \cellcolor[HTML]{d3d3d3}76.93\% \\
&\makecell[c]{Accuracy Improvement over~\cite{Minimax_approximation}} &+1.1\%(1.01$\times$) & +11.27\%(1.14$\times$) & +14.93\%(1.2$\times$) & +30.34\%(1.52$\times$) &+33.09\%(1.75$\times$) \\ 
 \bottomrule
\end{tabular}
\vspace{-4mm}
\label{tab:ablation}
\end{table*}

While PA doesn't always ensure superior accuracy—for instance, the baseline training (blue) outperforms PA in the $f_1 \circ g_2$ setup—it can enhance post-fine-tuning accuracy beyond the limits of higher-degree PAFs shown in \figref{fig:Progressive_Approximation}. Consequently, the \smartfhe framework divides training into smaller steps, proceeding with the best results between PA and baseline training, as depicted in \figref{fig:flow_chart}.

\subsection{Ablation Study}
\label{sec:ablation_study}

\subsubsection{Configuration Setup}
To investigate the effectiveness of different combinations of proposed techniques, we conduct an ablation study with results of VGG-19 (CiFar-10) ResNet-18 (ImageNet-1k) being presented in \tabref{tab:ablation}. We separate ReLU from MaxPooling to show the approximation effects of different types of non-polynomial operators. 

\squishlist
\item \textbf{Baseline + SS:} We take \cite{Lee2021PreciseAO, Minimax_approximation, cheon2020efficient} as the baseline. We obtain PAF coefficients through deterministic methods and uses one PAF to replace $sign(x)$ in all non-polynomial operators. Further, a fixed scale is chosen which falls into the category of Static Scaling. For a fair comparison, we still adopted Dynamic Scaling for the training baseline and converted it to a fixed scale after training. The final training accuracy before SS conversion is noted as ``baseline + DS".
\item \textbf{Baseline + ``technique" + DS:}  The final training accuracy using ``technique" before SS conversion.
\item \textbf{Baseline + ``technique" + SS:} The validation accuracy of FHE-deployable PAF-approximated model.
\squishend

\subsubsection{Impact of Techniques Combinations}

When replacing all non-polynomial operators, combining CT, PA, and AT yields the highest validation accuracy in PAF coefficients fine-tuning for ResNet-18 (Imagenet-1k). However, for VGG-19 (CiFar-10), there are outliers ($f_2\circ g_2$, and $f_1\circ g_2$) obtaining the highest validation accuracy using $CT$ only, as indicated by bolded values lying at different rows in \tabref{tab:ablation}. 

Further, when replacing ReLU only, the optimal accuracy point stems from a subset of proposed techniques, e.g. $\alpha=7$ obtains best training accuracy 68.4\% when using $PA$ while $f_2\circ g_3$ achieves 67.6\% when using $CT + PA$.
This variability inspired the development of step-wise \smartfhe framework, allowing us to test different combinations of proposed methods and select the most effective results for every step.

In short, the full combination of proposed techniques is prone to perform the best under complex tasks. While subsets of techniques might be sufficient for producing good accuracy for simpler tasks.

\subsubsection{Impact of Approximating MaxPooling}
The MaxPooling operator exhibits greater sensitivity to PAF replacement, evidenced by reduced accuracy in replacing both ReLU and MaxPooling than replacing ReLU only. This sensitivity arises because MaxPooling's single sliding window requires nested PAF calls, leading to the propagation and accumulation of approximation errors.

\begin{table*}[!t]\centering
\caption{\smartfhe V.S. \cite{Lee2021PreciseAO}. Results of two PAFs with better accuracy and latency are bolded.}
\vspace{-3mm}
\scriptsize
\begin{tabular}{cccccccc}\toprule

 VGG19 under CiFar-10, Original Accuracy $93.95$\%  & \multicolumn{5}{c}{\smartfhe} & Lee~\cite{Lee2021PreciseAO} \\
PAF format &$f_1\circ g_2$ & $f_2\circ g_2$& $f_2\circ g_3$ & $ \alpha=7$ & $f_1^2 \circ g_1^2$ & $\alpha=14$ \\   \hline 
 Validation Accuracy (Replace all non-polynomial) & 72.24\% &86.36\% &\textbf{91.05\%} & 91.82\% &\textbf{92.39\%} & 90.21\%\\ 
 Accuracy Improvement over ~\cite{Lee2021PreciseAO} & -19.97\% & -3.85\% & \textbf{+0.84\%} & +1.61\% &\textbf{+2.08\%} & 90.21\%\\ 
ReLU Latency on CPU (ms) & 3240.16 & 3510.82 & \textbf{4122.58} & 7113.35 & \textbf{6179.18} & 48278.84 (baseline) \\ 
Speedup over ~\cite{Lee2021PreciseAO} & 14.90$\times$ & 13.75$\times$ & \textbf{11.71$\times$} &6.79$\times$ & \textbf{7.81$\times$} & 1.0 \\ 
\bottomrule
\end{tabular} \\
\vspace{-3mm}
\label{tab:compare_vgg_cifar10}
\end{table*}

\insertWideFigureRatio{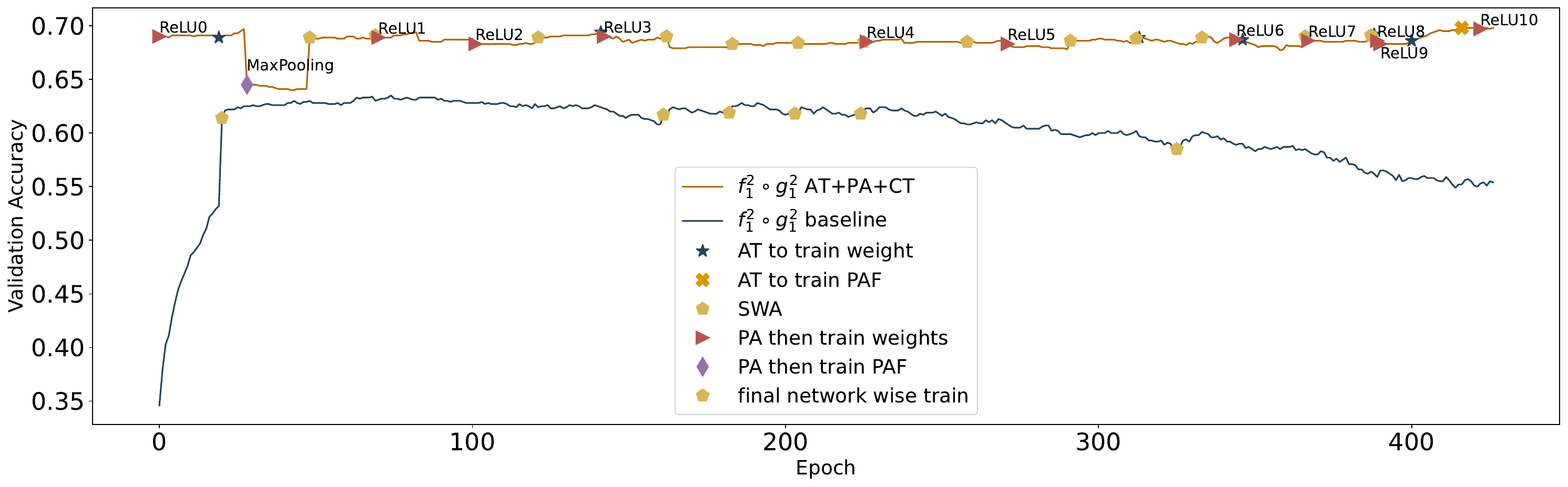}{Comparison of training curve (baseline v.s. \smartfhe)  ResNet-18 (ImageNet-1k) with ReLU approximated by PAF ($14$-degree $f_1^2 \circ g_1^2$)}{0.98}

\subsubsection{Impact of Dataset}
For VGG-19 under the CiFar-10 dataset, \smartfhe enhances the accuracy of prevailing Pareto-frontier by factors ranging from $1.01\times$ to $1.75\times$. This improvement jumps to $1.42\times$ up to $13.64\times$ for ResNet-18 under the ImageNet-1k dataset.
This distinction is attributed to the inherent complexity of ImageNet-1k, having $224\times224$ image size and 1k categories, compared to CiFar-10's $32\times32$ image size and 10 categories. 
For instance, replacing with $f_1^2\circ g_1^2$ results in a $0.55\%$ accuracy dip for VGG-19 on CiFar-10. In contrast, the same PAF leads to a significant $30.3\%$ decline for ResNet-18 on ImageNet-1k, highlighting the crucial role of task-tailored PAF selection in preserving accuracy.




\subsection{Comparison with SotA Implementations}
\subsubsection{Latency/Accuracy v.s. ~\cite{Lee2021PreciseAO}}
We also compare \smartfhe with the 27-degree PAF~\cite{Lee2021PreciseAO} widely used in prior FHE accelerators including F1\cite{f1} and BTS~\cite{BTS}.

For VGG-19 under CiFar-10 as shown in \tabref{tab:compare_vgg_cifar10}, \smartfhe identified two higher-accuracy PAFs with $7.81\times$ and $11.71\times$ speedup and $0.84\%$ and $2.08\%$ accuracy improvement. 

Considering ResNet-18 under ImageNet-1k, the PAF-approximated model using the 27-degree PAF demonstrates 69.3\% top-1 accuracy. In contrast, \smartfhe spotted the 14-degree PAF ($f_1^2\circ g_1^2$) that achieves $7.81\times$ speedup with 69.4\% PAF-approximated validation accuracy. Remarkably, this marginally surpasses original pretrained ResNet-18's accuracy 69.3\%, underscoring \smartfhe's efficacy.


\subsubsection{Accuracy v.s. ~\cite{Minimax_approximation, cheon2020efficient}}
We contrast \smartfhe with prevailing Pareto-frontier PAFs~\cite{Minimax_approximation} in \tabref{tab:ablation}. 
When all non-polynomial operators (ReLU and MaxPooling) were substituted, \smartfhe consistently achieves $1.42\times\sim13.64\times$ accuracy enhancement over~\cite{cheon2020efficient} on ResNet-18 (ImageNet-1k) and $1.01\times\sim1.75\times$ under VGG-19 (CiFar-10).

\section{Training Processing Deep Dive}

In this section, we perform a detailed comparison and analysis of the training curve using both the baseline training strategy and \smartfhe with the 14-degree PAF ($f_1^2 \circ g_1^2$)~\cite{Lee2021PreciseAO} in \tabref{tab:PAF_baseline}. Both baseline and \smartfhe train the PAF-approximated model step-by-step and leverage the same dropout and SWA strategy for a fair comparison.

Before the training, the baseline (blue in \figref{fig:convergence_curve}) replaces all non-polynomial operators with the same PAF, the coefficients of which are determined by regression algorithm~\cite{Lee2021PreciseAO, Minimax_approximation}. By contrast, \smartfhe only replaces the first non-polynomial layer with PAF followed by a Coefficient Tuning to reduce the accuracy drop, leaving other non-polynomial layers to be replaced progressively in the future steps, hence resulting in a 34.1\% initial accuracy improvement compared to the baseline as shown in \figref{fig:convergence_curve}.

During training, the accuracy of baseline drops after an initial boost,  Specifically, when the first training step ends ($E=20$ epochs), Stochastic Weights Averaging (SWA) is applied to smooth the weights update, recovering around 10\% accuracy demonstrated by the first yellow pentagon on the blue curve. However, every training step thereafter leads to worse accuracy as shown by the dropping trend of the blue curve, indicating a training failure because of no convergence guarantee under the SGD algorithm. 

On the contrary, \smartfhe replaces the ReLU progressively with post-CT PAFs, and validation accuracy increases for \smartfhe when training the PAF-approximated model under SGD.
During the middle of the training, each replacement of non-polynomial layer with PAF causes some accuracy degradation, which is further optimized back through both Stochastic Weight Averaging (SWA) and Alternate Training (AT), as shown in the climbing orange curve after each ReLU replacement (purple diamond) in \figref{fig:convergence_curve}. 

Moreover, AT enhances validation accuracy after replacing ReLU 0, 8, 9, and so on, as indicated by the ascending orange curve after applying AT (marked by an orange cross and dark star). However, AT shows no improvement for ReLU 3 and 6, consistent with findings in \tabref{tab:ablation}. This suggests that AT alone could help accuracy but doesn't guarantee accuracy enhancement for PAF-approximated models.

The detailed comparison illustrates both the invalidity of baseline training methods for PAF-approximated ML models and the effectiveness of \smartfhe.

\section{Related Work}
\label{sec:related}

ML inference is prone to a severe threat of private information leakage in the cloud. Fully Homomorphic Encryption (FHE)~\cite{homomorphicStandard} guarantees privacy by directly enabling computation on the encrypted data. 

FHE schemes do not support non-polynomial operators like ReLU or MaxPooling. Therefore, HEAX~\cite{heax}, Delphi~\cite{delphi}, Gazelle~\cite{Gazelle}, and Cheetah~\cite{cheetah} presented hybrid schemes consisting of FHE, Multi-Party Computation~\cite{deepsecure}, or  Garbled Circuits (GC)~\cite{GarbledCircuits}, where they relied on non-FHE scheme to process non-polynomial kernels. However, both MPC and GC are non-practical because of large-size packets in need of transferring among data sources and compute nodes.

Alternatively, \cite{Lee2021PreciseAO,cryptonets,Hesamifard2017CryptoDLDN,brutzkus2019low,CHE,Mishra2020DelphiAC} propose replacing all non-polynomial kernels with the same low-degree polynomial approximation and processing it in the FHE domain. However, a 27-degree PAF is used for low accuracy degradation, which consumes a large-portion of overall latency because of the long multiplication chain in FHE accelerators~\cite{f1, CraterLake, BTS}.

To further reduce the overhead of non-polynomial kernels, SAFENet~\cite{safenet}, and CryptoGCN~\cite{ran2022cryptogcn} proposed a finer granular replacement of non-polynomial operators with a lower degree, then they leverage ML training to figure out the parameters of each individual approximated polynomial. However, they suffer from significant accuracy degradation because of the training divergence in high-degree polynomials. Till now, there are no systematic training techniques for fine-tuning the model consisting of both linear operators and polynomial approximation functions. 

Recently, AESPA~\cite{park2022aespa} claims a quadratic approximation to replace ReLU with negligible accuracy degradation of VGG and ResNet models under CiFar-10/100 and TinyImageNet. However, the scheme does not show the method to approximate MaxPooling which is more sensitive and hard to approximate than ReLU as we quantified in \tabref{tab:ablation}. Further, the high accuracy preserving capability under TinyImageNet of quadratic does not guarantee low accuracy degradation under complex datasets like ImageNet, as we quantified in \secref{sec:ablation_study}.

Orthogonally, DeepReDuce~\cite{jha2021deepreduce} proposes reducing the number of ReLU, which could be leveraged together with SmartPAF to further reduce costs of privacy-preserving inference without hurting accuracy.


\section{Conclusion}
\label{sec:con}

This paper demonstrates that the training of ML models with non-polynomial operators replaced with Polynomial Approximated Functions (PAF) is a fundamentally different problem than the typical model training, such that typical training algorithms hardly converge and even lead to worse accuracy. 
This paper proposes four techniques and a training framework to address such a challenge:  
(1) Coefficient Tuning provides good initialization of PAF coefficients using profiled data distribution for higher post-replacement initial accuracy and faster convergence; (2) Progressive Approximation enables the PAF-approximated model training convergence under SGD algorithm through progressive layer-wise non-polynomial operators replacement and training; (3) Alternate Training separately trains PAF coefficients and layers except PAFs with different hyperparameters to avoid training interference; and (4) Dynamic Scale in training and Static Scale in post-training post-replacement inference under FHE to avoid value overflow. 
The order of applying techniques affects final accuracy, and \smartfhe framework is thus proposed to automatically apply the above techniques as well as dropout and SWA for accuracy improvement.
\smartfhe identifies the optimal Pareto-frontier in the latency-accuracy tradeoff space with $1.42\times\sim 13.64\times$ accuracy improvement and $6.79\times \sim 14.9\times$ speedup for ResNet-18 (ImageNet-1k). Further, \smartfhe spots the 14-degree PAF ($f_1^2\circ g_1^2$) that achieves the same 69.4\% post-replacement accuracy with $7.81\times$ latency speedup compared to 27-degree PAF obtained by minimax approximation. The validity and efficacy of \smartfhe are tested by multiple models and datasets. We believe that \smartfhe potentially opens a new paradigm of systematically augmenting PAF-approximated ML models for accurate and fast private inference.

\section{Acknowledge}
\label{sec:con}
We thank Guanghui Wang for mathematical formulation in \secref{sec:PAF_Statement}, Hanrui Wang for his valuable feedbacks to improve this paper. This work was supported in part by ACE, one of seven centers in JUMP 2.0, a Semiconductor Research Corporation (SRC) program sponsored by DARPA.

\nocite{langley00}

\bibliography{main}
\bibliographystyle{mlsys2024}

\appendix
\appendix

\section{Training Hyperparameters}

The following hyperparameters were used as the baseline training methodology to train the machine learning model.
\begin{table}[!h]
    \centering
    \small
    \caption{Baseline training parameters, where other layers include Convolution, Linear, and BatchNorm.}
    \label{tab:set_up}
    \vspace{-3mm}
\begin{tabular} {cc}
    \hline
    Configuration & Value \\
    \hline
    Replaced layer & ReLU\& MaxPooling \\
    Optimizer & Adam\\
    learning rate for PAF & 1e-4 \\
    learning rate for other layers & 1e-5 \\
    Weight decay for PAF & 0.01 \\
    Weight decay for other layers  & 0.1 \\
    BatchNorm Tracking & False \\ 
    Dropout & False \\ \hline
\end{tabular}
\end{table}

\section{Post-Training PAF Coefficients}
\label{sec:paf_form}
In this section, we present the specific coefficient values of polynomials with the highest validation accuracy shown in bold in \tabref{tab:ablation}. All proposed techniques including Coefficients Tuning (CT), Progressive Approximation (PA), and Alternate Training (AT) happen at the granularity of the layer and thus PAFs at different ReLU layers have different coefficients.

\subsection{$\alpha=7$}
\textbf{Mathematical Format}: The original format for the MiniMax approximated polynomial ($\alpha=7$)~\cite{Lee2021PreciseAO} is shown in \equref{equ:fun1}. Such a proposed polynomial $p_7(x)$ was used to approximate the $sign(x)$ function, which outputs $1$ when $x$ is positive and $-1$ when $x$ is negative. Then non-polynomial ReLU could be constructed using $\frac{x+x\cdot sign(x)}{2}$.

\begin{equation} \label{equ:fun1}
\begin{split}
p_7(x) &= p_{7,2}(x) \circ p_{7,1}(x) \\
p_{7,1}(x) = \sum_{i=0}^{7} a_i\times x^{i} & \ \ \ \ \ \ p_{7,2}(x) = \sum_{i=0}^{7} b_i\times x^{i} 
\end{split}
\end{equation}

The odd function nature of $sign(x)$ results in a negligible coefficient value of composite polynomial $p_7(x)$ in entries with an even degree of $x$. Such even-degree entries could be safely removed without the impact on the overall accuracy. Therefore, we remove all entries with even degrees to obtain $p_7(x)$ in the format shown in \equref{equ:fun2}.

\begin{equation} \label{equ:fun2}
\begin{split}
p_7(x) &= p_{7,2}^{odd\ only}(x) \circ p_{7,1}^{odd\ only}(x) \\
p_{7,1}^{odd\ only}(x) &= a_1  x + a_3  x^3 + a_5  x^5 + a_7  x^7 \\
p_{7,2}^{odd\ only}(x) &= b_1  x + b_3  x^3 + b_5  x^5 + b_7  x^7
\end{split}
\end{equation}

\begin{table}[!htp]\centering
\caption{$f_1 \circ g_2$ best coefficients list}\label{tab:best_f1_g2}
\scriptsize
\resizebox{\hsize}{!}{
\begin{tabular}{c|cc|ccc}\toprule
&\multicolumn{2}{c}{$f_2$ coefficients} &\multicolumn{3}{c}{$g_3$ coefficients} \\
layer id &$c_1$ &$c_3$ &$d_1$ &$d_3$ &$d_5$ \\\midrule
0  &3.064987659  &-4.359854698   &3.644091129  &-7.056697369  &4.412326813 \\
1  &2.939064741  &-3.989520550   &3.756805420  &-7.105865479  &4.209794998 \\
2  &2.962512255  &-4.095692158   &3.725888252  &-7.275540352  &4.892793179 \\
3  &2.996977568  &-4.153297901   &3.783520699  &-7.263069630  &4.682956696 \\
4  &2.898474693  &-4.044208527   &3.641639471  &-7.243083000  &4.771345139 \\
5  &2.895201445  &-3.905539751   &3.689141512  &-7.129144192  &4.736110687 \\
6  &3.018208981  &-4.113882542   &3.705801964  &-7.180747986  &4.518863201 \\
7  &2.848899364  &-3.874762058   &3.611979723  &-6.771905422  &4.524455547 \\
8  &3.008141994  &-4.087264061   &3.836204052  &-7.746193886  &4.919332504 \\
9  &2.968442440  &-3.986024141   &3.703149557  &-7.153123856  &4.776097775 \\
10  &2.900203228  &-3.924145937  &3.688660622  &-7.306476593  &4.663645267 \\
11  &2.782385111  &-3.684296608  &3.651248932  &-6.951449394  &4.715543270 \\
12  &2.958166838  &-3.980643034  &3.829906940  &-7.610838890  &4.719619274 \\
13  &2.811106443  &-3.719117880  &3.632898569  &-6.837011814  &4.688860893 \\
14  &2.911352396  &-3.886567831  &3.674616098  &-6.988801003  &4.670355797 \\
15  &2.796648502  &-3.706235886  &3.595447540  &-6.843948841  &4.560091972 \\
16  &3.042621136  &-3.979726553  &3.910200596  &-7.521365166  &4.733543873 \\
\bottomrule
\end{tabular}}
\end{table}

Under such polynomial format, coefficients used in PAF for replacing ReLU at different locations are shown in \tabref{tab:alpha7_layer0}.

\begin{table*}[!h]
    \centering
    \caption{Coefficients value for minimax composite polynomial ($\alpha=7$) used in PAF to replace all ReLU functions}
    \vspace{-3mm}
    \label{tab:alpha7_layer0}
    \scriptsize
    \begin{tabular}{cccccccc}\hline
    $a_1$ &$a_3$ &$a_5$ &$a_7$ &$b_1$ &$b_3$ &$b_5$ &$b_7$ \\
    7.304451 &-34.68258667 &59.85965347 &-31.87552261 &2.400856 &-2.631254435 &1.549126744 &-0.331172943 \\
    \hline
    \end{tabular}
\end{table*}

\begin{table*}[!htp]\centering
\caption{$f_1 \circ g_2$ Multiplication Depth Example}\label{tab:best_f1_g2}
\scriptsize
\vspace{-4mm}
\begin{tabular}{ccccccc}\hline
Multiplication Depth & 0& 1& 2& 3& 4& 5\\
Variables &$c_3$, $x$ & $c_3\cdot x$, $x^2$ & $c_3 \cdot x^3$, $y = f_1(x)$ & $d_5 \cdot y$, $y^2$ & $y^4$ & $d_5 \cdot y^5$\\
\hline
\end{tabular}
\vspace{-3mm}
\end{table*}

\subsection{$f_1^2 \circ g_1^2$}

The format of different building-block polynomials are shown in \equref{equ:f_g}.

\begin{equation}
\label{equ:f_g}
\begin{split}
f_1(x) = c_1\cdot x + c_3 \cdot x^3 \\
g_1(x) = d_1\cdot x + d_3 \cdot x^3 \\
f_2(x) = c_1\cdot x + c_3\cdot x^3 + c_5 \cdot x^5  \\
g_2(x) =  d_1\cdot x + d_3\cdot x^3 + d_5 \cdot x^5 \\
g_3(x) =  d_1\cdot x + d_3\cdot x^3 + d_5 \cdot x^5 + d_7\cdot x^7
\end{split}
\vspace{-3mm}
\end{equation}

$f_1^2 \circ g_1^2$ refers to combining both $f_1$ and $g_1$ into a composite polynomial in the sequence shown in \equref{equ:f1g1}, with PAF coefficients for at different ReLU layers shown in \tabref{tab:coef1g1}.

\begin{equation}
\label{equ:f1g1}
f_1^2 \circ g_1^2 (x) = g_1^1(g_1^0(f_1^1(f_1^0(x))))
\end{equation}


\begin{table*}[!htp]\centering
\caption{Coefficients of $f_1^2 \circ g_1^2$ at different layers}\label{tab:coef1g1}
\vspace{-3mm}
\scriptsize
\begin{tabular}{c|cccc|cccc}\toprule
&\multicolumn{4}{c}{$f_1^2$ coefficients} &\multicolumn{4}{c}{$g_1^2$ coefficients} \\
layer id &$c_1^0$ &$c_3^0$ &$c_1^1$ &$c_3^1$ &$d_1^0$ &$d_3^0$ &$d_1^1$ &$d_3^1$ \\\midrule
0 &2.736806631 &-3.864239931 &2.115309238 &-2.268822908 &2.239115477 &-2.424801588 &2.189934731 &-1.481475353 \\
1 &2.609737396 &-2.629375458 &2.115823507 &-1.854049206 &2.300836086 &-2.241225243 &2.231765747 &-1.455139399 \\
2 &2.572752714 &-2.620458364 &2.008517504 &-1.67325747 &2.017426491 &-1.779745221 &2.066540718 &-1.300397515 \\
3 &2.874353647 &-3.49595499 &2.073785543 &-1.72846055 &2.091589212 &-1.851963162 &2.141039133 &-1.372249603 \\
4 &2.588399172 &-3.086382866 &2.01845789 &-1.867060781 &1.999999881 &-1.845559597 &2.052644968 &-1.279196978 \\
5 &2.604569435 &-2.614924431 &1.93332684 &-1.466841698 &1.942190886 &-1.626866937 &2.10518527 &-1.243854761 \\
6 &2.510973692 &-2.517734289 &2.132683754 &-2.017316103 &2.235149622 &-2.204242945 &2.183528662 &-1.424280167 \\
7 &2.751836777 &-2.765525579 &2.021913052 &-1.521527886 &2.008341789 &-1.650658488 &2.125827074 &-1.320276856 \\
8 &2.517604351 &-2.519313574 &2.131887913 &-1.986418962 &2.247759819 &-2.206320763 &2.191907883 &-1.425198913 \\
9 &2.562408924 &-2.520729303 &2.110760212 &-1.814227581 &2.062101603 &-1.789000034 &2.126989841 &-1.338556409 \\
10 &2.437770844 &-2.398545027 &2.016869307 &-1.811605096 &2.103379965 &-1.996958494 &2.111694336 &-1.30810833 \\
11 &2.781474829 &-2.742717981 &2.02037096 &-1.498650432 &2.043134928 &-1.701895356 &2.140466452 &-1.345968127 \\
12 &2.483508587 &-2.447231293 &2.057531595 &-1.836180925 &2.189022541 &-2.110060215 &2.162631512 &-1.370931029 \\
13 &2.787295341 &-2.709958792 &2.00928688 &-1.456294537 &2.007162809 &-1.627877712 &2.114115715 &-1.327487946 \\
14 &2.674963474 &-2.604590893 &2.028381109 &-1.637359142 &2.129605532 &-1.939982772 &2.159248829 &-1.392939448 \\
15 &2.731667519 &-2.661221027 &2.026224852 &-1.519181132 &2.036108494 &-1.692675114 &2.118255377 &-1.338307023 \\
16 & 2.670770168	&-2.607930183	&2.119180441&	-1.756756186&	2.236502171&	-2.061469316&	2.230870724&	-1.45818007\\
\bottomrule
\end{tabular}
\end{table*}

Similarly, coefficients value of $f_2 \circ g_3$,  $f_2 \circ g_2$ and $f_1 \circ g_2$ are shown in \tabref{tab:best_f2_g3}, \tabref{tab:best_f2_g2} and \tabref{tab:best_f1_g2}, separately.





\begin{table*}[!htp]\centering
\caption{$f_2 \circ g_3$ best coefficients list}\label{tab:best_f2_g3}
\vspace{-3mm}
\scriptsize
\begin{tabular}{c|ccc|cccc}\toprule
&\multicolumn{3}{c}{$f_2$ coefficients} &\multicolumn{4}{c}{$g_3$ coefficients} \\
layer id &$c_1$ &$c_3$ &$c_5$ &$d_1$ &$d_3$ &$d_5$ &$d_7$ \\\midrule
0 &3.487593412 &-6.971315384 &2.381806374 &4.736026287 &-16.16058159 &25.20542908 &-13.1174 \\
1 &3.484929323 &-7.034649372 &3.685389519 &4.983552456 &-17.01627541 &25.34817886 &-12.4504 \\
2 &3.312547922 &-6.849102974 &3.659186125 &4.616300583 &-15.70791912 &25.24704933 &-13.7765 \\
3 &3.42953968 &-7.291306973 &3.949234486 &4.785545349 &-16.25030518 &25.22435379 &-13.1702 \\
4 &3.550015688 &-7.992001534 &3.389156818 &4.644083023 &-15.87583256 &25.47412872 &-13.8047 \\
5 &3.484149933 &-7.679964066 &3.130941153 &4.65158844 &-15.79552174 &25.19073868 &-13.6172 \\
6 &1.875 &-1.25 &0.375 &4.481445313 &-16.18847656 &25.01367188 &-12.5586 \\
7 &3.137469292 &-6.013744831 &2.900674343 &4.600552082 &-15.5252409 &24.95741463 &-13.7303 \\
8 &3.355214119 &-5.68600893 &1.215050697 &4.856618881 &-16.73614693 &25.50185585 &-12.7147 \\
9 &3.605870724 &-9.147006989 &6.160003185 &4.596205711 &-15.64334202 &25.45436478 &-14.1617 \\
10 &3.669521809 &-8.906849861 &5.65577507 &4.712775707 &-16.15146828 &25.63137817 &-13.6679 \\
11 &3.432019472 &-8.035040855 &4.964941978 &4.565317631 &-15.44346809 &25.10269928 &-13.9918 \\
12 &3.677670956 &-8.38080883 &4.933722496 &4.846800804 &-16.69511223 &25.66197395 &-13.0236 \\
13 &3.383493662 &-8.223423958 &5.385590076 &4.52063942 &-15.19449425 &24.9539814 &-14.2344 \\
14 &3.32148385 &-7.110795498 &4.014864445 &4.572896957 &-15.55243587 &25.26078415 &-14.0067 \\
15 &3.381628513 &-7.793000221 &4.806651115 &4.586762428 &-15.50544167 &25.14218521 &-14.0126 \\
16 &3.627621889 &-8.305987358 &5.061814785 &4.829498291 &-16.53964996 &25.57732391 &-13.1699 \\
\bottomrule
\end{tabular}
\end{table*}



\begin{table*}[!htp]\centering
\caption{$f_2 \circ g_2$ best coefficients list}\label{tab:best_f2_g2}
\vspace{-3mm}
\scriptsize
\begin{tabular}{c|ccc|ccc}\toprule
&\multicolumn{3}{c}{$f_2$ coefficients} & \multicolumn{3}{c}{$g_3$ coefficients} \\
layer id &$c_1$ &$c_3$ &$c_5$ &$d_1$ &$d_3$ &$d_5$ \\\midrule
0  &3.632708073  &-8.879578590  &4.333632946  &3.700465441  &-7.351731300  &5.071476460 \\
1  &3.412810802  &-7.752333164  &4.516210556  &3.855783939  &-7.789761543  &5.177268505 \\
2  &3.355527401  &-8.588312149  &5.618574142  &3.640014887  &-7.615984440  &5.668038368 \\
3  &3.533123493  &-9.278223038  &6.205972672  &3.779361486  &-7.770857811  &5.565216064 \\
4  &1.875000000  &-1.250000000  &0.375000000  &3.255859375  &-5.964843750  &3.707031250 \\
5  &3.421332598  &-9.231142044  &6.353975773  &3.687772274  &-7.753697395  &5.787805080 \\
6  &3.494106293  &-8.028047562  &3.792766333  &3.851673841  &-8.117405891  &5.920250893 \\
7  &3.236023188  &-7.844894886  &4.858978271  &3.662446976  &-7.398378849  &5.480692863 \\
8  &3.308430910  &-7.289185524  &3.084533691  &3.766145468  &-8.078896523  &5.651748657 \\
9  &3.438756227  &-9.819555283  &7.128154278  &3.620871305  &-7.664072514  &5.793798447 \\
10  &3.470819712  &-9.487674713  &6.564511299  &3.746651173  &-8.130080223  &6.042979240 \\
11  &3.344857931  &-8.513930321  &5.686520100  &3.717740774  &-7.314604759  &5.406781673 \\
12  &3.561307669  &-9.413117409  &6.282663822  &3.941442251  &-8.642221451  &6.365680695 \\
13  &3.235330582  &-8.009678841  &5.256969452  &3.645334482  &-7.250671864  &5.429522514 \\
14  &3.269543648  &-7.355520248  &4.257196426  &3.702267408  &-7.359237194  &5.368722439 \\
15  &3.318752050  &-8.203745842  &5.435956478  &3.630973339  &-7.331366062  &5.393109322 \\
16  &3.595479012  &-9.167343140  &6.192716122  &3.955091715  &-8.303151131  &6.023469925 \\
\bottomrule
\end{tabular}
\end{table*}



\section{Multiplication Depth Analysis}

CKKS (Cheon-Kim-Kim-Song) is a leveled homomorphic encryption scheme capable of evaluating $L$-level arithmetic circuits without the need for bootstrapping. Here, $L$ signifies the depth of the arithmetic circuit that can be computed, which is contingent upon the parameters defining the homomorphic context. Every Rescaling or modulus reduction decreases one level to reduce the noise of the multiplication operation. 

In this context, the ``multiplication depth" of a PAF refers to the number of levels reduced during its evaluation. The multiplication depth is also decided by the highest degree term in PAF. To minimize level consumption in PAF multiplication, contemporary methodologies leverage the exponentiation by squaring strategy (See \figref{fig:multiplication_depth}). For instance, for a polynomial featuring a highest degree term of $a\cdot x^n$, the required depth is computed as $\left \lceil{\log_2(n+1)}\right \rceil$. When dealing with composite polynomials, the overall depth necessitated is the aggregate of the depths required for each constituent sub-polynomial.
    
Taking $f_1\circ g_2= g_2(f_1(x))$ as an example. The multiplication depth of each intermediate result is shown in \tabref{equ:f_g} with the illustration shown in \figref{fig:multiplication_depth}.

\begin{equation}
    \label{equ:f_g}
    \begin{split}
    y = f_1(x) = c_1\cdot x + c_3 \cdot x^3 \\
    g_2(y) = d_1\cdot y + d_3 \cdot y^3 + d_5 \cdot y^5
    \end{split}
\end{equation}

\insertFigureRatio{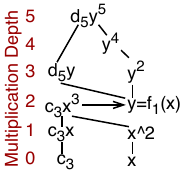}{Illustration of multiplication depth for $f_1\circ g_2$.}{0.4}%

\end{document}